\documentclass[10pt,twocolumn,twoside]{article}
\usepackage{amssymb}
\usepackage{amsmath}
\usepackage[dvips]{graphicx}
\usepackage[latin1]{inputenc}

\providecommand{\dsum}{\sum}

\begin{document}

\title{Electrostatic internal energy using the method of images}
\author{Rodolfo A. Diaz\thanks{%
radiazs@unal.edu.co}, William J. Herrera\thanks{%
jherreraw@unal.edu.co}, J. Virgilio Niño\thanks{
jvninoc@unal.edu.co} \\
Departamento de Física. Universidad Nacional de Colombia. Bogotá, Colombia.}
\date{}
\maketitle

\begin{abstract}
\emph{For several configurations of charges in the presence of conductors,
the method of images permits us to obtain some observables associated with
such a configuration by replacing the conductors with some image charges.
However, simple inspection shows that the potential energy associated with
both systems does not coincide. Nevertheless, it can be shown that for a
system of a grounded or neutral conductor and a distribution of charges
outside, the external potential energy associated with the real charge
distribution embedded in the field generated by the set of image charges is
twice the value of the internal potential energy associated with the
original system. This assertion is valid for any size and shape of the
conductor, and regardless of the configuration of images required. In
addition, even in the case in which the conductor is not grounded nor
neutral, it is still possible to calculate the internal potential energy of
the original configuration through the method of images. These results show
that the method of images could also be useful for calculations of the
internal potential energy of the original system. }


\end{abstract}

\vspace{3mm}

The method of images is a useful tool to calculate electrostatic fields,
Green functions and forces that conductors of certain shapes exert over some
charges \cite{Berkeley}-\cite{Griffiths}. However, the internal energy of
the configuration is seldom treated by this method \cite{Pomer}, and
applications are restricted to very special cases \cite{Griffiths}. Let us
call \textbf{system} \textbf{A} (the real system)\textbf{\ }the one
consisting of a conductor and certain distribution of charges outside it,
and \textbf{system} \textbf{B} (the virtual system) the one consisting of
the distribution of charges plus the set of image charges, see Fig. \ref%
{fig:AB}. The reason that prevents us from calculating the energy of system
A based on system B, is that the integral $\int \mathbf{E}^{2}dV$ is not the
same for both configurations, because in the region inside the conductor the
electric field is different in each system. For an arbitrary size and shape
of the conductor there is no evident symmetry to connect the energies of
both configurations. Remarkably, despite the lack of symmetry, it can be
demonstrated that for system A with the conductor grounded or neutral, the
external potential energy associated with the real distribution of charges
in the presence of the set of images is twice the value of the internal
potential energy associated with system A. This fact is true for an
arbitrary size and shape of the conductor, and whatever the configuration of
images is. Therefore, the method of images could also be utilized for
calculations of the internal potential energy of the original system.
Finally, the result could be extended for the case in which the conductor is
not neccesarily grounded nor neutral.

\begin{figure}[tbh]
\begin{center}
\includegraphics[width=6.5cm]{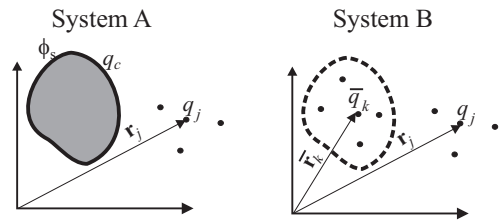}
\end{center}
\caption{\textit{System A is defined as the set composed of the conductor
and the distribution of charges $\left\{ q_{j}\right\} $ outside the
conductor (left). System B consists of the distribution of charges $\left\{
q_{j}\right\} $ plus the set of image charges $\left\{ \bar{q}_{k}\right\} $%
, (right)}}
\label{fig:AB}
\end{figure}

\section{General approach}

We want to find the internal potential energy associated with a system
consisting of a point charge $q$ in the presence of a conductor (system A);
we shall start from the expression

\begin{equation}
U_{int}^{\left( A\right) }=\frac{1}{2}\int \rho \left( \mathbf{r}\right) \
\phi \left( \mathbf{r}\right) \ dV,  \label{Uint}
\end{equation}%
where $\rho \left( \mathbf{r}\right) ,\ \phi \left( \mathbf{r}\right) $
denote the charge density and potential, respectively. It is well known that
Ec. (\ref{Uint}) leads to divergences when point particles are present
because of the inclusion of self energy terms \cite{Griffiths}. Of course,
the self energy term can be removed to get only finite results. Extracting
the self energy and taking into account that the integral only contributes
in regions where charge is present, we get%
\begin{equation}
U_{int}^{\left( A\right) }=\frac{1}{2}q_{c}\phi _{s}+\frac{1}{2}q\phi
_{A}\left( \mathbf{r}_{0}\right) ,  \label{UA}
\end{equation}%
where $\phi _{s}$ is the potential on the surface of the conductor, $\mathbf{%
r}_{0}$ describes the location of the point charge, $q_{c}$ is the net
charge of the conductor, and $\phi _{A}\left( \mathbf{r}_{0}\right) $ is the
electric potential at $\mathbf{r}_{0}$ due to all sources (excluding $q$
itself i.e. removing the divergence). From the method of images \cite%
{Berkeley}-\cite{Griffiths}, the electric potential outside the conductor is
equivalent to the electric potential generated by system B (the charge $q$
plus the set of images). In particular, the electric potential generated by
the set of image charges at the point $\mathbf{r}_{0}$ is given by\footnote{%
Once again, the self potential generated by the point charge at $\mathbf{r}%
_{0}$ is extracted.}%
\begin{equation}
\phi _{B}\left( \mathbf{r}_{0}\right) =\sum_{k=1}^{N}\frac{K_{c}\bar{q}_{k}}{%
\left\vert \mathbf{\bar{r}}_{k}-\mathbf{r}_{0}\right\vert }=\phi _{A}\left( 
\mathbf{r}_{0}\right) \ ,  \label{fiAr0}
\end{equation}%
where $\left( \bar{q}_{k},\mathbf{\bar{r}}_{k}\right) $ denotes the set of
image charges and their positions, and $K_{c}\equiv \left( 4\pi \varepsilon
_{0}\right) ^{-1}$ in SI units. Replacing (\ref{fiAr0}) into (\ref{UA}) we
find%
\begin{equation}
U_{int}^{\left( A\right) }=\frac{1}{2}q_{c}\phi _{s}+\frac{1}{2}\left[
\sum_{k=1}^{N}\frac{K_{c}\bar{q}_{k}q}{\left\vert \mathbf{\bar{r}}_{k}-%
\mathbf{r}_{0}\right\vert }\right] \ .  \label{UintB}
\end{equation}%
From Gauss's law, it can be seen that the net charge $q_{c}$ on the surface
of the conductor is the algebraic sum of the image charges. Similarly, the
potential on the surface of the conductor is the potential generated by
system B in any point of such a surface; hence we get%
\begin{equation}
q_{c}=\sum_{k=1}^{N}\bar{q}_{k}\ \ ;\ \ \phi _{s}=\frac{K_{c}q}{\left\vert 
\mathbf{r}_{0}-\mathbf{r}_{s}\right\vert }+\sum_{k=1}^{N}\frac{K_{c}\bar{q}%
_{k}}{\left\vert \mathbf{\bar{r}}_{k}-\mathbf{r}_{s}\right\vert }\ ,
\label{qfis}
\end{equation}%
where $\mathbf{r}_{s}$ is the position of any point on the surface of the
conductor. Replacing (\ref{qfis}) into (\ref{UintB}), we get the internal
energy of system A in terms of the components of system B%
\begin{eqnarray}
U_{int}^{\left( A\right) } &=&\frac{1}{2}\left( \sum_{k=1}^{N}\bar{q}%
_{k}\right) \left( \frac{K_{c}q}{\left\vert \mathbf{r}_{0}-\mathbf{r}%
_{s}\right\vert }+\sum_{m=1}^{N}\frac{K_{c}\bar{q}_{m}}{\left\vert \mathbf{%
\bar{r}}_{m}-\mathbf{r}_{s}\right\vert }\right)   \notag \\
&&+\frac{1}{2}\left[ \sum_{k=1}^{N}\frac{K_{c}\bar{q}_{k}q}{\left\vert 
\mathbf{\bar{r}}_{k}-\mathbf{r}_{0}\right\vert }\right] .  \label{Uint2}
\end{eqnarray}%
The term in square brackets in Eq. (\ref{Uint2}) is the potential energy
associated with the point charge $q$ in the presence of the set of images,
which we shall denote as $U_{q}^{\left( B\right) }$ getting%
\begin{equation}
U_{int}^{\left( A\right) }=\frac{1}{2}q_{c}\phi _{s}+\frac{1}{2}%
U_{q}^{\left( B\right) }.  \label{UA-Uq}
\end{equation}%
Equation (\ref{UA-Uq}) is valid for an arbitrary form of the conductor, and
allows us to find the internal potential energy of the real system, based on
the potential energy associated with the point charge $q$ in the presence of
the image charges. Result (\ref{UA-Uq}) can be further generalized by using
the principle of superposition. So if instead of a point charge we have a
distribution of $M$ charges denoted by $\left\{ q_{j}\right\} $, Eq. (\ref%
{UA-Uq})$\ $becomes%
\begin{eqnarray}
U_{int}^{\left( A\right) } &=&\frac{1}{2}q_{c}\phi _{s}+\frac{1}{2}%
U_{ext}^{\left( B\right) }\ ,  \notag \\
q_{c} &=&\sum_{k=1}^{N}\bar{q}_{k}\ \ ;\ \ \phi _{s}=\sum_{j=1}^{M}\frac{%
K_{c}q_{j}}{\left\vert \mathbf{r}_{j}-\mathbf{r}_{s}\right\vert }%
+\sum_{m=1}^{N}\frac{K_{c}\bar{q}_{m}}{\left\vert \mathbf{\bar{r}}_{m}-%
\mathbf{r}_{s}\right\vert }  \notag \\
U_{ext}^{\left( B\right) } &\equiv &\sum_{j=1}^{M}q_{j}\sum_{k=1}^{N}\frac{%
K_{c}\bar{q}_{k}}{\left\vert \mathbf{\bar{r}}_{k}-\mathbf{r}_{j}\right\vert }%
=\sum_{j=1}^{M}q_{j}\bar{\phi}\left( \mathbf{r}_{j}\right) \ ,
\label{gen result}
\end{eqnarray}%
where $\bar{\phi}\left( \mathbf{r}_{j}\right) $ is the potential generated
by the set of images in the location $\mathbf{r}_{j}$ of the charge $q_{j}$.
Thus, $U_{ext}^{\left( B\right) }$ represents the external potential energy
associated with the distribution of real charges when they are immersed in
the field generated by the images. Of course, the distribution (and perhaps
the images) could also be continuous; in that case the sums become
integrals. In particular, if the conductor is grounded or neutral we find
that \footnote{%
Notwithstanding, the result is not necessarily the same for null charge as
for the zero potential because the set of images is not in general the same
in both cases.}%
\begin{equation}
U_{int}^{\left( A\right) }=\frac{1}{2}U_{ext}^{\left( B\right) }\ .
\label{U grounded}
\end{equation}

At this point it would be convenient to discuss briefly about the difference
between external and internal potential energies \cite{Finn}. First, we
should precise the system of particles for which we define the concepts of
external and internal. Once the system is defined, the internal potential
energy is that associated with the internal forces and corresponds to the
work necessary to ensemble the system starting from the particles far away
from each other. On the other hand, the external potential energy is that
associated with the external forces, and corresponds to the work necessary
to bring the system as a whole from infinity to their final positions%
\footnote{%
It is assumed that the system is already ensembled and that it moves as a
rigid body, in order to ensure that the internal energy is not changing in
the process. The initial point could be another point of reference different
from infinity.}, immersed in the field of forces generated by all sources
outside the system. In our case, $U_{ext}^{\left( B\right) }$ represents the
external potential energy associated with the system of real charges in
which the external forces are provided by the system of images. Therefore, $%
U_{ext}^{\left( B\right) }$ is the work necessary to bring the distribution $%
\left\{ q_{j}\right\} $ as a whole, from infinity to their final positions
in the presence of the image charges. On the other hand, $U_{int}^{\left(
A\right) \text{ }}$ represents the work necessary to ensemble the system
consisting of the conductor and the real charges which is the observable we
are interested in. Note that internal and external potential energies are
conceptually and operationally different.

\section{Simple applications}

We shall work on two types of scenarios (a) The conductor is isolated (so
the net charge $q_{c}$ is fixed), and (b) The potential on the surface of
the conductor is fixed (e.g. grounded or connected to a battery).

\textbf{Example 1:\ }The most typical problem treated by the method of
images is the one of a point charge and an infinite grounded conducting
plane. For many purposes, such a system is equivalent to replacing the
conductor by the appropiate image charge, forming a physical dipole. Ref. 
\cite{Griffiths}, shows that the energy stored in system A, is only half of
the energy stored in system B. It can be seen from either of the following
arguments, (a) The space can be divided into two halves separated by the
conductor. For system B the integral $\int \mathbf{E}^{2}dV$ gives identical
contributions in both halves, while in system A only one of those halves
contributes to the energy. (b) Calculating the work necessary to bring $q$
from infinity. In system A we only work on $q$ since the redistribution of
charges in the conductor costs nothing because those charges are moving on
an equipotential. By contrast, if we assemble system B by bringing both
charges simultaneously, we could work on both of them symmetrically
resulting in a work twice as great. This solution agrees with (\ref{U
grounded}), and can be guessed by arguments of symmetry. However, Eq. (\ref%
{U grounded}) holds far beyond of this example.

\textbf{Example 2}: A point charge $q\ $at $\mathbf{r}_{0}\ $in the presence
of an isolated, charged conductor with net charge $q_{c}$. We are interested
in calculating the external work to bring $q$ from infinity to $\mathbf{r}%
_{0}$. The net charge is invariant during the process and the internal
energy of the system at the beginning and at the end of the process are
obtained by applying (\ref{UintB})%
\begin{equation*}
U_{int}^{\left( A,i\right) }=\frac{1}{2}q_{c}\phi _{s}^{i}\ ;\
U_{int}^{\left( A,f\right) }=\frac{1}{2}q_{c}\phi _{s}^{f}+\frac{1}{2}\left[
\dsum\limits_{k}\frac{K_{c}\bar{q}_{k}^{\left( f\right) }}{\left\vert 
\mathbf{r}_{0}-\mathbf{\bar{r}}_{k}^{\left( f\right) }\right\vert }q\right]
\ ,
\end{equation*}%
where $\phi _{s}^{i},\phi _{s}^{f}$ are the potentials on the surface of the
conductor at the beginning and at the end of the process respectively, the
set of images $\left\{ \bar{q}_{k}^{\left( f\right) },\mathbf{\bar{r}}%
_{k}^{\left( f\right) }\right\} $ are those that fit the potential of the
conductor at the end of the process (with $q$ lying at $\mathbf{r}_{0}$). We
have assumed that the set of images is localized all over the process to
ensure that the potential energy associated with the point charge is zero
when it is located at infinity. The external work to bring $q$ from infinity
to $\mathbf{r}_{0}$, is the change of internal energy%
\begin{equation}
W_{ext}=\Delta U_{int}^{\left( A\right) }=\frac{1}{2}q_{c}(\phi
_{s}^{f}-\phi _{s}^{i})+\frac{1}{2}\left[ \dsum\limits_{k}\frac{K_{c}\bar{q}%
_{k}^{\left( f\right) }}{\left\vert \mathbf{r}_{0}-\mathbf{\bar{r}}%
_{k}^{\left( f\right) }\right\vert }q\right] \ ,  \label{Wext isolated}
\end{equation}%
the values of $\phi _{s}^{i}$ and $\phi _{s}^{f}$ can be obtained from Eq. (%
\ref{qfis}) by utilizing the configuration of images at the beginning and at
the end of the process respectively\footnote{$q_{c}$ is given in the
problem. But for consistency we could check whether it is obtained by using
the set of images at the beginning or at the end of the process in Eq. (\ref%
{qfis}), since $q_{c}$ is invariant during the process.}. It is clear that
the values of $\phi _{s}^{f},\ \phi _{s}^{i}$ depend on the geometry of the
conductor. Notice however that if the net charge is null, $W_{ext}$ becomes
independent of those potentials and the result has the same form as that of
the grounded conductor (see Eqs. \ref{gen result}, \ref{U grounded})
regardless of the geometry of the conductor.

\textbf{Example 3}: Point charge $q$ at $\mathbf{r}_{0}$ with a conductor
connected to a battery that keeps it at a fixed potential $V$. In the
process of bringing $q\ $from infinity to $\mathbf{r}_{0}$, the battery must
supply a charge $\Delta Q$ to the conductor to maintain a constant voltage,
so that applying (\ref{UintB}) at the beginning and at the end and making
the difference, we obtain the change in the internal energy

\begin{equation*}
\Delta U_{int}^{\left( A\right) }=\frac{V}{2}\Delta Q+\frac{1}{2}\left[
\dsum\limits_{k}\frac{K_{c}\bar{q}_{k}^{\left( f\right) }}{\left\vert 
\mathbf{r}_{0}-\mathbf{\bar{r}}_{k}^{\left( f\right) }\right\vert }q\right]
\ .
\end{equation*}%
Again, we have assumed that the set of images is localized all over the
process. Let us denote $q_{c}^{\left( i\right) },q_{c}^{\left( f\right) }$
as the total charges on the surface of the conductor at the beginning and at
the end of the process respectively. Using Eq. (\ref{qfis}), the change in
internal energy becomes

\begin{eqnarray}
\Delta U_{int}^{\left( A\right) } &=&\frac{V}{2}\left[ \left( \dsum_{k}\bar{q%
}_{k}^{\left( f\right) }\right) -\left( \dsum_{m}\bar{q}_{m}^{\left(
i\right) }\right) \right]  \notag \\
&&+\frac{1}{2}\left[ \dsum\limits_{k}\frac{K_{c}\bar{q}_{k}^{\left( f\right)
}}{\left\vert \mathbf{r}_{0}-\mathbf{\bar{r}}_{k}^{\left( f\right)
}\right\vert }q\right] \ .  \label{dU battery}
\end{eqnarray}%
This change of internal energy is equal to the net external work on the
system, which can be separated in the work done by the battery to supply
charge to the conductor plus the work due to the external force acting on $q$

\begin{equation*}
\Delta U_{int}^{\left( A\right) }=W_{ext}=W_{batt}+W_{F_{ext}}\ .
\end{equation*}%
Further, the work done by the battery is%
\begin{equation}
W_{batt}=V\ \Delta Q=V\left[ \left( \dsum_{k}\bar{q}_{k}^{\left( f\right)
}\right) -\left( \dsum_{m}\bar{q}_{m}^{\left( i\right) }\right) \right] \ ,
\label{Wbatt}
\end{equation}%
so that $W_{F_{ext}}$ reads

\begin{eqnarray}
W_{F_{ext}} &=&\frac{V}{2}\left[ \left( \dsum_{m}\bar{q}_{m}^{\left(
i\right) }\right) -\left( \dsum_{k}\bar{q}_{k}^{\left( f\right) }\right) %
\right]  \notag \\
&&+\frac{1}{2}\left[ \dsum\limits_{k}\frac{K_{c}\bar{q}_{k}^{\left( f\right)
}}{\left\vert \mathbf{r}_{0}-\mathbf{\bar{r}}_{k}^{\left( f\right)
}\right\vert }q\right] \ ,  \label{WFext}
\end{eqnarray}%
the grounded conductor is a special case with $V=0$.

Examples 2, 3 are valid for any size and shape of the conductor. Let us
apply these results for a spherical conductor of radius $R$, with the origin
at the center of the sphere.

\begin{figure}[tbh]
\begin{center}
\includegraphics[width=6.2cm]{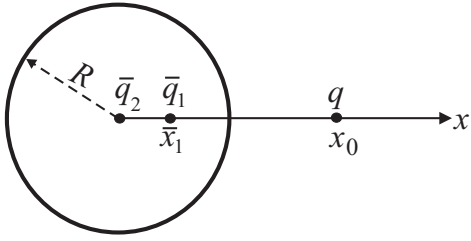}
\end{center}
\caption{\textit{A point charge $q\ $\textit{in the presence of a conducting
sphere of radius }$R$\textit{. The charges }$\bar{q}_{1},\bar{q}_{2}$ are
the set of images, and they acquire different values and positions according
to the case studied. However $\bar{q}_{2}$ is always at the origin and the
set\textit{\ }$\bar{q}_{1},\bar{q}_{2},q$ lies on the $X-$axis.}}
\label{fig:sphere}
\end{figure}

\textbf{Example 4:} The sphere is connected to a battery that keeps its
potential $V\ $constant. The structure of the set of images is well known
from the literature, (see for instance sections 2.2, 2.4 of \cite{Jackson}).
In the notation of Fig. \ref{fig:sphere}, for the charge $q$ at certain
position $x_{0}$, the structure of the images is given by%
\begin{equation}
\bar{q}_{1}=-\frac{qR}{x_{0}}\ ;\ \bar{x}_{1}=\frac{R^{2}}{x_{0}}\ ;\ \bar{q}%
_{2}=\frac{VR}{K_{c}}\ ;\ \bar{x}_{2}=0\ ,  \label{images2}
\end{equation}%
the images at the beginning are obtained by taking $x_{0}\rightarrow \infty $%
, and at the end we just assume that the position is $x_{0}$; using (\ref%
{qfis}) we find%
\begin{equation}
q_{c}^{\left( i\right) }=\bar{q}_{1}^{\left( i\right) }+\bar{q}_{2}^{\left(
i\right) }=\frac{VR}{K_{c}}\ ;\ q_{c}^{\left( f\right) }=\bar{q}_{1}^{\left(
f\right) }+\bar{q}_{2}^{\left( f\right) }=-\frac{qR}{x_{0}}+\frac{VR}{K_{c}}
\label{qci}
\end{equation}%
replacing (\ref{images2}, \ref{qci}) into (\ref{Wbatt}, \ref{WFext}) we find%
\begin{eqnarray}
W_{batt} &=&-\frac{qRV}{x_{0}}  \notag \\
\ W_{F_{ext}} &=&\frac{qRV}{x_{0}}-\frac{1}{2}\frac{K_{c}q^{2}R}{\left(
x_{0}^{2}-R^{2}\right) }  \notag \\
&=&\frac{K_{c}\bar{q}_{2}q}{x_{0}}+\frac{1}{2}\frac{K_{c}q\bar{q}%
_{1}^{\left( f\right) }}{\left( x_{0}-\bar{x}_{1}^{\left( f\right) }\right) }%
\ .  \label{WFext2}
\end{eqnarray}%
the grounded sphere is obtained by setting $V=0$ (or $\bar{q}_{2}=0$). By
examining the last line of Eq. (\ref{WFext2}) we see that the first term of $%
W_{F_{ext}}$ is equivalent to the work to bring the charge $q$ in the
presence of the image $\bar{q}_{2}$ (which is invariant during the process).
It is worth emphasizing that in this term the factor $1/2$ is not present
because the charge $\bar{q}_{2}$ is stationary and constant in magnitude in
the process of bringing $q$, so that the external force necessary to bring
the charge is always of the form $K_{c}\bar{q}_{2}q/r^{2}$ in the direction
of motion. By contrast, in the second term the factor $1/2$ is present and
such a term is equivalent to half of the work required to carry the $q\ $%
charge in the presence of the image $\bar{q}_{1}\ $if such an image always
remained in its final position, and with its final magnitude. This factor
arises from the fact that during the process of bringing $q$, the image
charge $\bar{q}_{1}$ must change its position and magnitude, in order to
retain its role as image charge.

\textbf{Example 5}: The sphere is isolated with net charge $q_{c}$. Again,
the structure of images required appears in the literature (see for instance
sections 2.2, 2.3 of \cite{Jackson}), and in the notation of Fig. \ref%
{fig:sphere} they read%
\begin{equation}
\bar{q}_{1}=-\frac{qR}{x_{0}}\ ;\ \bar{x}_{1}=\frac{R^{2}}{x_{0}}\ ;\ \bar{q}%
_{2}=q_{c}-\bar{q}_{1}=q_{c}+\frac{qR}{x_{0}}\ ;\ \bar{x}_{2}=0\ ,
\label{images}
\end{equation}%
the potential on the surface of the conductor when the charge lies in its
final position, owes to the charge $\bar{q}_{2}$ only, since the potential
generated by $\bar{q}_{1}$ and $q$ cancel each other by construction. On the
other hand, the potential on the surface of the conductor when $q$ lies at
infinity, is clearly of the form $K_{c}q_{c}/R$, and using (\ref{images})
the potentials on the surface of the conductor at the beginning and at the
end of the process read 
\begin{equation}
\phi _{s}^{f}=\frac{K_{c}\bar{q}_{2}}{R}=\frac{K_{c}\left( q_{c}+\frac{qR}{%
x_{0}}\right) }{R}\ \ ;\ \ \phi _{s}^{i}\ =\frac{K_{c}q_{c}}{R}\ ,
\label{potentials}
\end{equation}%
replacing expressions (\ref{images}, \ref{potentials}) into (\ref{Wext
isolated}) we get%
\begin{equation*}
W_{ext}=K_{c}q\left\{ \frac{q_{c}}{x_{0}}+\frac{qR}{2}\left[ \frac{1}{%
x_{0}^{2}}-\frac{1}{\left( x_{0}^{2}-R^{2}\right) }\right] \right\} \ .
\end{equation*}%
In this case, there is no work on the conductor similar to the case of a
grounded sphere ($V=0$ in Eq. \ref{WFext2}). In particular, in the scenario
with a neutral conductor i.e. $q_{c}=0$, the work necessary to bring the
charge is larger than in the case of a grounded sphere because a second
image charge located at the centre and of the same sign of $q\ $must be
added, hence it leads to a repulsion that requires the external work to be
increased.

\begin{figure}[tbh]
\begin{center}
\includegraphics[width=6.0cm]{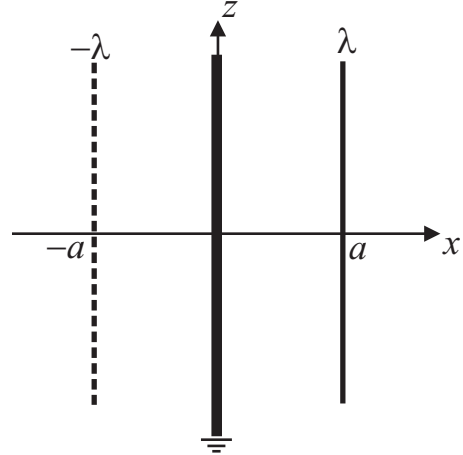}
\end{center}
\caption{\textit{An infinite wire with constant linear charge density $%
\protect\lambda $,\ in front of a grounded conducting plane, and its
corresponding image.}}
\label{fig:wire}
\end{figure}

\textbf{Example 6}: Let us consider an infinite wire with uniform linear
density $\lambda $, lying at a distance $a$ on the right-side of an infinite
grounded conducting plane, see Fig. \ref{fig:wire}. In this case the image
consists of another infinite wire of linear density $-\lambda $ on the
left-side of the plane. In this example the real and image distributions are
continuous. The electric potentials in a point $\mathbf{r=}x\mathbf{\hat{%
\imath}}+y\mathbf{\hat{\jmath}+}z\mathbf{\hat{k}}$ ($x>0$) due to the wire
and its image are given by

\begin{eqnarray}
\Phi (\mathbf{r}) &=&2K_{c}\lambda \ln \sqrt{(x-a)^{2}+y^{2}}+C_{1},  \notag
\\
\bar{\Phi}(\mathbf{r}) &=&-2K_{c}\lambda \ln \sqrt{(x+a)^{2}+y^{2}}+C_{2}.
\label{Pot_wire_image}
\end{eqnarray}

To ensure that the potential over the plane is null ($YZ$ plane) we should
choose the arbitrary constants as $C_{1}=-C_{2}=C$ so that the total
potential on the right-side of the plane becomes

\begin{equation*}
\Phi _{T}(\mathbf{r})=2K_{c}\lambda \ln \sqrt{\frac{(x-a)^{2}+y^{2}}{%
(x+a)^{2}+y^{2}}},
\end{equation*}%
which satisfies the condition $\Phi _{T}(0,y,z)=0$. Now we intend to
calculate the electrostatic internal energy of the system A, for which we
use Eq. (\ref{gen result}) with $\phi _{s}=0$%
\begin{equation}
U_{int}^{\left( A\right) }=\frac{1}{2}U_{ext}^{\left( B\right) }\ \ \ ;\ \ \
U_{ext}^{\left( B\right) }\equiv \int \bar{\Phi}(a\mathbf{,}0\mathbf{,}0)\ dq
\label{Uintext}
\end{equation}%
where the external potential energy per unit length associated with the real
wire in the field generated by the virtual one (system B) can be written as

\begin{equation*}
\frac{U_{ext}^{(B)}}{L}=\bar{\Phi}(a\mathbf{,}0\mathbf{,}0)\lambda ,
\end{equation*}%
with $L$ being the length of a piece of the wire with density $\lambda $,
from Eq. (\ref{Pot_wire_image}) we get

\begin{equation*}
\frac{U_{ext}^{(B)}}{L}=-2K_{c}\lambda ^{2}\ln \left( 2a\right) -C\lambda .
\end{equation*}

Using Eq. (\ref{Uintext}) the potential energy per unit length of the wire
in the presence of the grounded conducting plane can be written as

\begin{equation*}
\frac{U_{int}^{\left( A\right) }}{L}=-K_{c}\lambda ^{2}\ln \left( 2a\right) -%
\frac{C\lambda }{2}\ ,
\end{equation*}%
and the external work per unit length to carry the wire (in the presence of
the conductor) from a distance $a_{i}$ to a distance $a_{f}$ reads

\begin{equation*}
\frac{W_{ext}^{^{_{i\rightarrow f}}}}{L}=-K_{c}\lambda ^{2}\ln \left( \frac{%
a_{i}}{a_{f}}\right) .
\end{equation*}

\section{Conclusions}

By using the method of images we have found a general equation that permits
us to calculate the internal potential energy associated with a
configuration consisting of a conductor and a distribution of charges
outside the conductor. Such an equation shows that the work necessary to
bring the distribution of charges in the presence of a grounded or neutral
conductor is just half of the external potential energy associated with the
real charges in the presence of the field produced by the image charges.
This result is general and does not depend on any special symmetry of the
system. Finally, even for no grounded nor neutral conductors the internal
energy of the system can be obtained through the images. In particular, we
studied the case of isolated conductors and conductors connected to a
battery.

In summary, we have used the method of images to calculate the internal
energy associated with some electrostatic configurations. Commom textbooks
do not use the method of images for calculations of electrostatic internal
energies, except for the case of the point charge in front of an infinite
grounded conducting plane because of its very high symmetry. Finally, Ref. 
\cite{Pomer} treats a similar problem but its approach requires the full
Green formalism and is not very easy for practical calculations. By
contrast, the formalism presented here only requires the knowledge of the
method of images and basic concepts of potential energy. Besides, practical
calculations with our approach are very easy.


\begin{thebibliography}{9}
\bibitem{Berkeley} E. Purcell \textquotedblleft \emph{Electricity and
Magnetism}\textquotedblright\ 2nd Ed., McGrawHill (1985); R. Feynman, R.
Leighton, M. Sands, \textquotedblleft \emph{The Feynman Lectures on Physics
Vol. II}\textquotedblright\ Addison-Wesley Publishing Co. (1964).

\bibitem{Jackson} J. D. Jackson \textquotedblleft \emph{Classical
Electrodynamics}\textquotedblright\ 3rd Ed., John Wiley \& Sons (1998).

\bibitem{Griffiths} David J. Griffiths \textquotedblleft \emph{Introduction
to Electrodynamics}\textquotedblright\ 3rd Ed., Prentice Hall (1999).
Section 3.2.3

\bibitem{Pomer} F. Pomer, \textquotedblleft \emph{Electric energy and forces
in the presence of boundary conditions}\textquotedblright\, Am. J. Phys. 
\textbf{56}, 262 (1988).

\bibitem{Finn} Marcelo Alonso, Edward Finn \textquotedblleft \emph{%
Fundamental University Physics\ Vol I: Mechanics}\textquotedblright\ Addison
Wesley (1967) Chap. 9.
\end{thebibliography}
\end{document}